\documentclass[english,twocolumn,prx]{revtex4-1}
\usepackage[T1]{fontenc}
\usepackage[latin9]{inputenc}
\usepackage{mathrsfs}
\usepackage{amsmath}
\usepackage{amssymb}
\usepackage{graphicx}

\makeatletter
\usepackage{tikz}
\newcommand{\sq}[1][0.18]{\tikz[baseline=-.3pt,line width=.7 pt/0.18*#1,scale=#1,cap=rect]{
\draw (0,0)--(0,1)--(0.5,1)--(0.5,0)--(1,0)--(1,1)--(1.5,1)--(1.5,0)--(2,0)--(2,1)}}
\newcommand{\saw}[1][0.18]{\tikz[baseline=-.3pt,line width=.7 pt/0.18*#1,scale=#1,cap=rect]{
\draw (0,1)--(0,0)--(1,1)--(1,0)--(2,1)--(2,0)}}
\newcommand{\tri}[1][0.18]{\tikz[baseline=-.3pt,line width=.7 pt/0.18*#1,scale=#1,cap=rect]{
\draw (0,1)--(0.5,0)--(1,1)--(1.5,0)--(2,1)--(2.5,0)}}
\newcommand{\sawr}[1][0.18]{\tikz[baseline=-.3pt,line width=.7 pt/0.18*#1,scale=#1,cap=rect]{
\draw (0,0)--(0,1)--(1,0)--(1,1)--(2,0)--(2,1)}}
\DeclareMathOperator{\sinc}{sinc}

\makeatother

\usepackage{babel}
\begin{document}

\title{Coherent spin-wave processor of stored optical pulses}

\author{Mateusz Mazelanik}
\email{mateusz.mazelanik@fuw.edu.pl}

\affiliation{Faculty of Physics, University of Warsaw, Pasteura 5, 02-093 Warsaw,
Poland}

\author{Micha\l{} Parniak}
\email{michal.parniak@fuw.edu.pl}

\affiliation{Faculty of Physics, University of Warsaw, Pasteura 5, 02-093 Warsaw,
Poland}

\author{Adam Leszczy\'{n}ski}

\affiliation{Faculty of Physics, University of Warsaw, Pasteura 5, 02-093 Warsaw,
Poland}

\author{Micha\l{} Lipka}

\affiliation{Faculty of Physics, University of Warsaw, Pasteura 5, 02-093 Warsaw,
Poland}

\author{Wojciech Wasilewski}

\affiliation{Faculty of Physics, University of Warsaw, Pasteura 5, 02-093 Warsaw,
Poland}
\begin{abstract}
A device being a pinnacle of development of an optical quantum memory
should combine the capabilities of storage, inter-communication and
processing of stored information. In particular, the ability to capture
a train of optical pulses, interfere them in an arbitrary way and
finally perform on-demand release would in a loose sense realize an
optical analogue of a Turing Machine. Here we demonstrate the operation
of an optical quantum memory being able to store optical pulses in
the form of collective spin-wave excitations in a multi-dimensional
wavevector space. During storage, we perform complex beamsplitter
operations and demonstrate a variety of protocol implemented as the
processing stage, including interfering a pair of spin-wave modes
with 95\% visibility. By engineering the phase-matching at the readout
stage we realize the on-demand retrieval. The highly multimode structure
of the presented quantum memory lends itself both to enhancing classical
optical telecommunication as well as parallel processing of optical
qubits at the single-photon level. 
\end{abstract}
\maketitle

\section{Introduction}

As optical quantum memory technologies are becoming more mature, the
range of of their applications increases. The basic memories operating
in a single temporal and spatial mode can store only one optical pulse
and interfere it with second pulse only during light-atom coupling
\citep{Reim2012}. Such memories, based either on Raman scattering
or electromagnetically induced transparency (EIT), can achieve high
efficiencies \citep{Vernaz-Gris2018}, but offer very limited capacity
as multiplexing is limited by the number of atomic magnetic sublevels
employed \citep{Wang2011,Lee2014,Vernaz-Gris2018,Xu2013}. While a
single atomic ensemble may be split into an array to offer parallel
storage of light \citep{Lan2009,Pu2017}, such scheme hinders manipulations
within the memory as communication between memory cells must be inherently
light-based. It is thus highly desirable to independently store many
optical pulses within the same group of atoms. Such a multiplexing
scheme may utilize either the spatial \citep{Parniak2017,Ding2013}
or temporal degree of freedom \citep{Gundogan2013,Kutluer2017,Tiranov2016}.
In the latter case considered in the context of the atomic-ensemble
based quantum memories the Gradient Echo Memory (GEM) \citep{Nunn2008,Hosseini2009,hosseini2011high,Sparkes2013,Albrecht2015a}
scheme stands out as an efficient way to engineer the phase-matching
at readout stage to achieve mode-selective storage and retrieval.
Similar feature is inherently offered by the atomic frequency comb
(AFC) memories based on ensembles of ions in solids \citep{Kutluer2017,Gundogan2013,Sinclair2014}
thanks to their large inhomogeneous broadening. In the spatial degree
of freedom atomic ensembles allow storage of light in many angular-emission
modes through spin-wave wavevector multiplexing \citep{Dai2012,Parniak2017}.
These schemes allow storage of hundreds of optical modes, also when
used with non-classical states of light.

Manipulation of stored optical pulses, however, remains a substantial
challenge, both from technical and fundamental points of view. The
AFC memory has been demonstrated to allow pre-programmed interference
of two stored pulses with a single-output port \citep{Sinclair2017,Saglamyurek2014},
and within the GEM scheme a beamsplitter operation between a pre-selected
stored pulses and an input pulse has been realized \citep{Campbell2012,Campbell2014}.
These schemes also allow basic spectral and temporal manipulations
of stored light. More work is needed however to reach the regime of
efficient and arbitrary manipulations of stored light. In particular,
the ac-Stark shift caused by an additional light field has been proposed
as a versatile way to realize the GEM scheme \citep{Sparkes2010}.
Recent theoretical proposals went beyond the simple gradient shape
and suggested to engineer the stored spin-wave shape to realize Kapitsa-Dirac
diffraction \citep{Hetet2015} or a quantum memory protected with
a disordered password \citep{Su2017}. Finally, a recent experiment
used the ac-Stark shift to realize a spin-wave beamsplitter at the
single-excitation level demonstrating Hong-Ou-Mandel interference
for stored light \citep{Parniak2018}.

Here we present the first realization of ac-Stark-based spin-wave
universal multiport interferometric processor (SUMIP) and join the
advantages of the transverse-wavevector and temporal multiplexing
to realize a variety of operations on the stored coherent states of
light. The previously untackled regime of complex light patterns used
to engineer spin waves is explored, which allows us to tap into the
full three-dimensional potential of the wavevector-multiplexed optical
quantum memory. We show that thanks to engineering of the spatial
profile of ac-Stark modulation the stored pulses may be processed,
interfered and conditionally retrieved. The scheme features both reprogrammable
reordering and interference of pulses within the multiple-input, multiple-output
paradigm, essential to realize true unitary operations. In the paper
we first introduce the protocol by deriving its theoretical principles
and realizing a scheme reminiscent of the Gradient Echo Memory \citep{Hosseini2009}.
Next, we realize a series of programmable beamsplitting experiments
in spatial and temporal degrees of freedom. High-visibility interference
of a pair of modes is demonstrated. Finally, we propose potential
further applications and give technical details of the experiment
and light-atom propagation simulations involved. 

\section{Operation of the light-atom interface}

The atomic optical memory based on an elongated ensemble of Rb-87
atoms employs a strong control field $\mathcal{{E}}_{C}$ to map a
weak signal field $\mathcal{{E}}_{in}$ onto the atomic coherence
$\rho_{gh}$ between the two meta-stable ground states, for which
we take $|g\rangle\rightarrow$ $F=1,\ m_{F}=-1$, and $|h\rangle\rightarrow$
$F=2,\ m_{F}=1$ (see Fig. 1c for the atomic level scheme). In the
experiment the atoms are first optically pumped to the $|g\rangle$
state and control and signal fields operate with opposite circular
polarizations. Typically we use 300 ns long pulses for storage and
retrieval of atomic coherence.

While the interaction is well characterized by a set of coupled Maxwell-Bloch
equations (see Appendix A), first we rather choose to describe the
atom-light coupling qualitatively. In particular, Fig. 1b illustrates
the geometry in which the coupling and signal fields co-propagate
through an elongated atomic ensemble. Assuming that the coupling beam
diameter is significantly larger than the transverse size of the ensemble,
we may actually solve the coupled equations within the first order
in the coupling strength and obtain a simple result by which a signal
$\mathcal{E}(k_{x},k_{y})$ couples to an atomic coherence 
\begin{multline*}
\rho_{gh}(K_{x},K_{y},K_{z})\propto\\
\mathcal{{E}}_{in}(k_{x}=K_{x},k_{y}=K_{y})\exp(i\Delta_{0}t)\delta(K_{z0}-K_{z}),
\end{multline*}
where $K_{z0}=\sqrt{\omega^{2}/c^{2}-k_{x}^{2}-k_{y}^{2}}-\omega_{C}/c$,
$\delta$ is a Dirac delta function and $\omega$ and $\omega_{C}$
are frequencies of signal and coupling fields, respectively, and $c$
is the speed of light. For $k_{x}=k_{y}=0$ the longitudinal wavevector
simplifies to a constant component $K_{z0}=c/\Delta_{0}\approx0.14\ \mathrm{rad}\ \mathrm{m}\mathrm{m}^{-1}$,
where $\Delta_{0}\approx2\pi\times6.8\ \mathrm{{GHz}}$ is the nominal
frequency splitting between levels $|g\rangle$ and $|h\rangle$.
To exclude these trivial dependencies from further consideration we
will define the stored spin-wave excitation as 
\begin{multline*}
S(K_{x},K_{y},K_{z})=\rho_{gh}(K_{x},K_{y},K_{z})\ast\\
\mathscr{F}[\sqrt{N(x,y,z)}](K_{x},K_{y},K_{z}-K_{z0})\exp(-i\Delta_{0}t),
\end{multline*}
where $\mathscr{F}$ stands for the Fourier transform in the spatial
domain, $N(x,y,z)$ is the atom number density and $\ast$ denotes
convolution (here in the wavevector space). Importantly, after mapping
the optical field we obtain a spin-wave excitation with $K_{z}=0$
in terms of $S$. 

The process of reverse mapping or retrieval driven by the same coupling
field occurs in a symmetric way. Essentially, an atomic spin-wave
excitation will be mapped onto an optical field proportional to $S$
in terms of the transverse wavevector dependence only if $K_{z}=0$.
This requirement arises due to the phase-matching condition. In particular,
the allowed spread in the $K_{z}$ space is inversely proportional
to the atomic cloud length $\sigma_{z}$ and most importantly spin
waves with large $K_{z}$ component ($K_{z}\sigma_{z}\gg1$) will
remain stored in the memory. This remains true unless we change the
frame of reference significantly by selecting much different $K_{x}$,
$K_{y}$, as the actual phase matching is satisfied on an ellipsoid
in a $K$ space rather than a plane. Its curvature will depend on
the particular geometrical configuration. Here we will remain within
the regime where we may use the phase-matching planar approximation
to consider which spin waves are retrievable.

\section{Spin-wave manipulation with the ac-Stark effect}

\begin{figure}
\includegraphics[width=1\columnwidth]{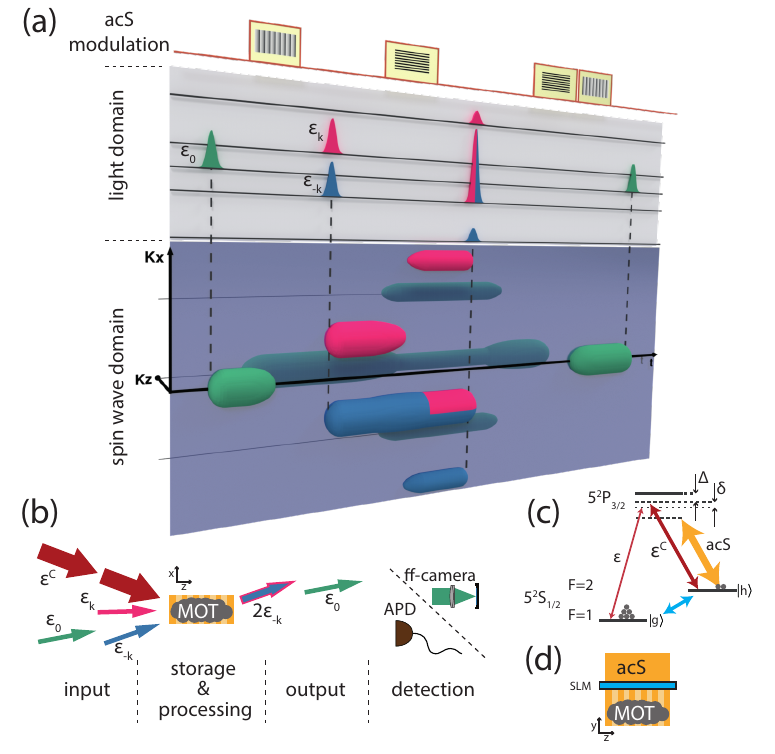}\caption{Spin-wave processing with ac-Stark modulation. (a) Simplified simulation
of exemplary protocol implemented on SUMIP, presenting the light and
spin-wave domain transformations. By manipulating the spin wave using
spatially varying ac-Stark (acS) shift three pulses are stored, processed
and released in reversed order. In the three-dimensional plot, the
colored blobs correspond to a spin-wave density exceeding a critical
value. We coloring is selected to correspond to a light pulse from
which the spin wave originates. (b) Pulses are sent to SUMIP in different
time-space modes, then ac-Stark manipulation is applied and at the
end pulses are released. Released pulses can be detected either on
Avalanche Photodiode (APD) or sCMOS camera situated in the far-field
(ff) for temporal or angular resolution, respectively. (c) Rubidium-87
energy level configuration utilized for storing and processing coherent
light pulses. (d) Projection of the panel (b) on perpendicular plane
exposing 2-d $(x,z)$ ac-Stark modulation capabilities using Spatial
Light Modulator (SLM).}
\end{figure}

As discussed above, only a limited space, or more precisely a thin
three-dimensional volume around $K_{z}=0$ plane in the wavevector
space may be populated by spin waves by means of Raman interaction.
To manipulate the spin waves within and beyond this volume we use
an additional far off-resonant beam {[}marked in Fig. 1(c) as acS{]}
that induces an additional differential ac-Stark shift between levels
$|g\rangle$ and $|h\rangle$ of $\Delta_{\mathrm{acS}}$ that adds
to $\Delta_{0}$. The ac-Stark beam propagating along the $y$-direction
is $z$-polarized and red-detuned by approx. 1 GHz from the $|h\rangle\to|e\rangle$
transition and is inducing $\Delta_{\mathrm{acS}}\sim$1 MHz ac-Stark
shift with $\sim100$ mW beam power. This shift causes the atomic
coherence $\rho_{gh}$, and thus the spin wave, to accumulate an additional
phase $\varphi_{\mathrm{acS}}=\Delta_{\mathrm{acS}}T$ over the interaction
time $T$. Typically we use ac-Stark pulses of approx. $T\sim2\ \mu\mathrm{s}$
duration. By spatially shaping the ac-Stark beam intensity $I_{\mathrm{acS}}(x,z)$
we induce a spatially-dependent phase shift $\varphi_{\mathrm{acS}}(x,z)\propto I_{\mathrm{acS}}(x,y)$,
which due to the geometry of the experiment is limited to two dimensions
{[}see Figure 1(b){]}. Any spin wave is then reshaped as: 
\begin{multline}
S(K_{x},K_{y},K_{z})=\\
\iint_{(X,Z)}\mathscr{F}[\exp(i\varphi_{\mathrm{acS}}(x,z))](k_{x},k_{z})\\
S(K_{x}+k_{x},K_{y},K_{z}+k_{z})\mathrm{d}k_{x}\mathrm{d}k_{z}.
\end{multline}
 A basic example is an ac-Stark analogue of the GEM, in which a phase
shift linear in $z$ $(\varphi_{\mathrm{acS}}=\beta z$) induced by
a magnetic field gradient shifts the spin wave in the $K_{z}$ direction
by $\beta$.

\begin{figure*}
\includegraphics[width=1\textwidth]{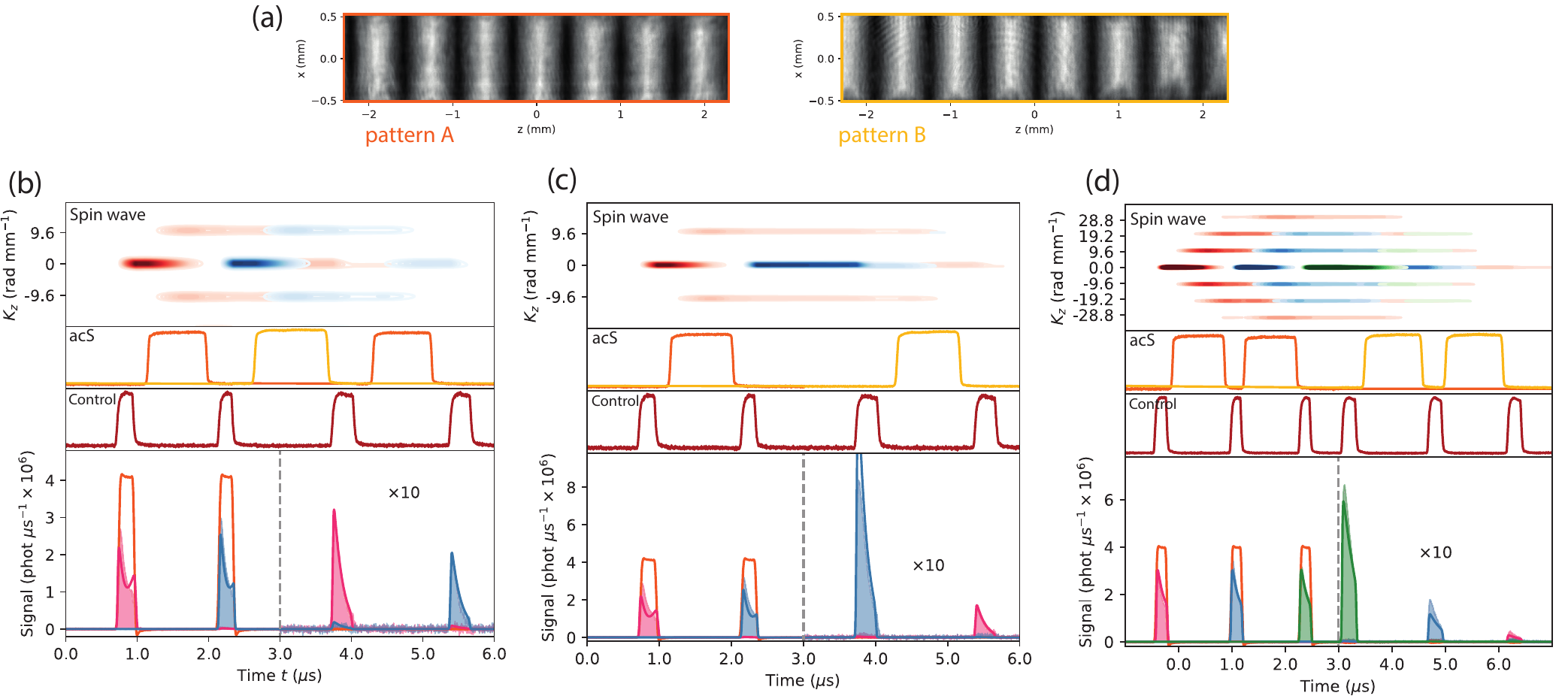}

\caption{Operation of the ac-Stark Echo Memory. Panel (a): ac-Stark triangle-wave
($\tri$) modulation patterns for storage (pattern A) retrieval (pattern
B) of subsequent pulses. Pattern B is a shifted by a half of a period
version of pattern A with the same amplitude. Storage and retrieval
of two coherent optical pulses in the \textit{last-in, first-out}
(LIFO) as well as \textit{first-in, first-out} (FIFO) configurations
{[}panels (b) and (c), respectively{]} and the LIFO configuration
for storage of three pulses {[}panel (d){]}.}
\end{figure*}

Here we work with optical modes characterized by small wavevector
(angular) spread and therefore the spin waves are well-localized in
the wavevector space. Discrete mode transformation in such a space
are most conveniently performed by a spatially periodic ac-Stark modulation.
Taking a spatial period $2\pi/k_{\mathrm{acS}}$ of the modulation
in the form $\varphi_{\mathrm{acS}}(x,z)=\varphi_{\mathrm{acS}}^{\mathrm{periodic}}(\mathbf{k}_{\mathrm{acS}}\cdot(x,z))$
we may express the spin-wave transformation using Fourier series as:

\begin{equation}
S(\mathbf{K})\xrightarrow{\varphi_{\mathrm{acS}}(x,z)}\sum_{n=-\infty}^{+\infty}c_{n}S(\mathbf{K}+n\mathbf{k}_{\mathrm{acS}}),
\end{equation}
with Fourier coefficients $c_{n}$ defined as: 
\begin{equation}
c_{n}=\frac{1}{2\pi}\int_{0}^{2\pi}\exp(i\varphi_{\mathrm{acS}}^{\mathrm{periodic}}(\xi)-in\xi)\mathrm{d\xi.}
\end{equation}

Figure 1(a) presents a simplified simulation (assuming perfect write-in
and read-out) of a protocol operation within this paradigm. In this
exemplary protocol three pulses $\mathcal{E}_{0}$, $\mathcal{E}_{-k}$,
$\mathcal{E}_{k}$ with transverse optical wavevector component $k_{x}$
equal 0, $-k$ and $k$, respectively, are stored, processed and released
from the memory. When the first pulse is mapped to the ensemble, the
created spin wave is phase-modulated using a sawtooth-shaped $\saw$
(periodic) modulation $\varphi_{\mathrm{acS}}(x,z)=\varphi_{\mathrm{acS}}^{\saw[0.08]}(k_{\saw[0.08]}z)$.
Such a modulation with amplitude equal $2\pi$ shifts the spin wave
in the $K_{z}$ direction by $k_{\saw[0.08]}$, making the spin wave
unreadable, as $k_{\saw[0.08]}\sigma_{z}\gg1$. Then, two pulses with
$k_{x}$ components separated by $2k$ arriving at the same time are
written to the ensemble. Next, square-shaped $\sq$ modulation $\varphi_{\mathrm{acS}}(x,z)=\varphi_{\mathrm{acS}}^{\sq[0.08]}(k_{\sq[0.08]}x)$
with $k_{\sq[0.08]}=2k$ is applied. The amplitude and phase of this
modulation are chosen so the first three Fourier coefficients are
in following relations $c_{-1}^{\sq[0.08]}=c_{0}^{\sq[0.08]}=-c_{1}^{\sq[0.08]}$
. This way the two pulses with non-zero $k_{x}$ component are combined
in a way reminiscent of a two mode beam-splitter transformation \citep{Parniak2018}
resulting in constructive interference in the mode with $k_{x}=-k$.
Note that the spin wave component corresponding to the first pulse
is transformed as well (split), but because of shifted $K_{x}$ and
most importantly $K_{z}$ component it does not take part in the interference.
At this stage the first readout is performed during which all readable
(zero $K_{z}$) spin wave components are converted to light pulses.
To read out the first pulse, all the previous transformations are
undone by applying versions of previous modulations shifted by half
a period in reversed order. At the very end of the protocol the first
pulse is retrieved. 

\section{Reconfigurable ac-Stark Echo Memory}

For the experimental demonstration we begin by moving the spin waves
outside the zero $K_{z}$ to allow storage of subsequent incoming
optical pulses. This configuration, most reminiscent of the GEM, here
operates best with a triangle-shaped $\tri$ grating (pattern A),
which can be conventionally written in a closed form: 
\begin{equation}
\varphi_{\mathrm{acS}}^{\tri[0.08]}(\xi)=\mathcal{A}^{\tri[0.08]}\left|2\left(\frac{\xi}{2\pi}-\left\lfloor \frac{\xi}{2\pi}+\frac{1}{2}\right\rfloor \right)\right|,
\end{equation}
 with $\mathbf{k}_{\tri[0.08]}=9.6\mathrm{mm^{-1}}\hat{e}_{z}$ for
which most essentially the zeroth order $c_{0}^{\tri[0.08]}\propto|\sinc(\mathcal{A}/2)|$
disappears periodically with modulation strength $\mathcal{A}^{\tri[0.08]}$
(with period equal $2\pi$) except for$\mathcal{A}^{\tri[0.08]}=0$.
With this scheme we may thus apply a grating with $\mathcal{A}^{\tri[0.08]}=2\pi$
and remove the pulse from the $K_{z}=0$ plane. Due to the periodicity
of $c_{0}^{\tri[0.08]}$ in the modulation strength $\mathcal{A}^{\tri[0.08]}$,
if a subsequent pulse is stored, the first and any previous pulse
remains phase-mismatched at consecutive grating operations with amplitude
$\mathcal{A}^{\tri[0.08]}=2\pi$. To retrieve the pulses we apply
a pattern with the same amplitude shifted by a half of period in the
spatial domain (pattern B), that restores the spin waves to the $K_{z}=0$
plane.

The scheme lends itself to both \textit{first-in, first-out} (FIFO)
and \textit{last-in, first-out} (LIFO) operation, as shown in Fig.
2. For the FIFO operation on two pulses, after storage of a second
pulse we apply a shifted pattern B to simultaneously transfer the
first pulse back to the $K_{z}=0$ plane and phase-mismatch the second
pulse. After first retrieval operation the phase matching is restored
for the second pulse with pattern A. 

The efficiency of our memory is currently limited by the optical depth
of the ensemble as well as available coupling power. By comparing
the intensity of light at the input and output of the memory we obtain
write-in efficiency for the first pulse of about 59\% and 44\% for
the second pulse. For immediate retrieval (as for the second pulse
in LIFO scheme) we achieve 35\% efficiency of retrieval, while net
storage and retrieval efficiency is equal $44\%\times35\%=15\%$.
For the pulses that are manipulated the efficiency is diminished by
dephasing due to the ac-Stark light intensity inhomogenities \citep{Leszczynski2018}
(see Appendix B for details).

\section{Programmable beamsplitting of stored optical pulses}

\begin{figure}
\includegraphics[width=1\columnwidth]{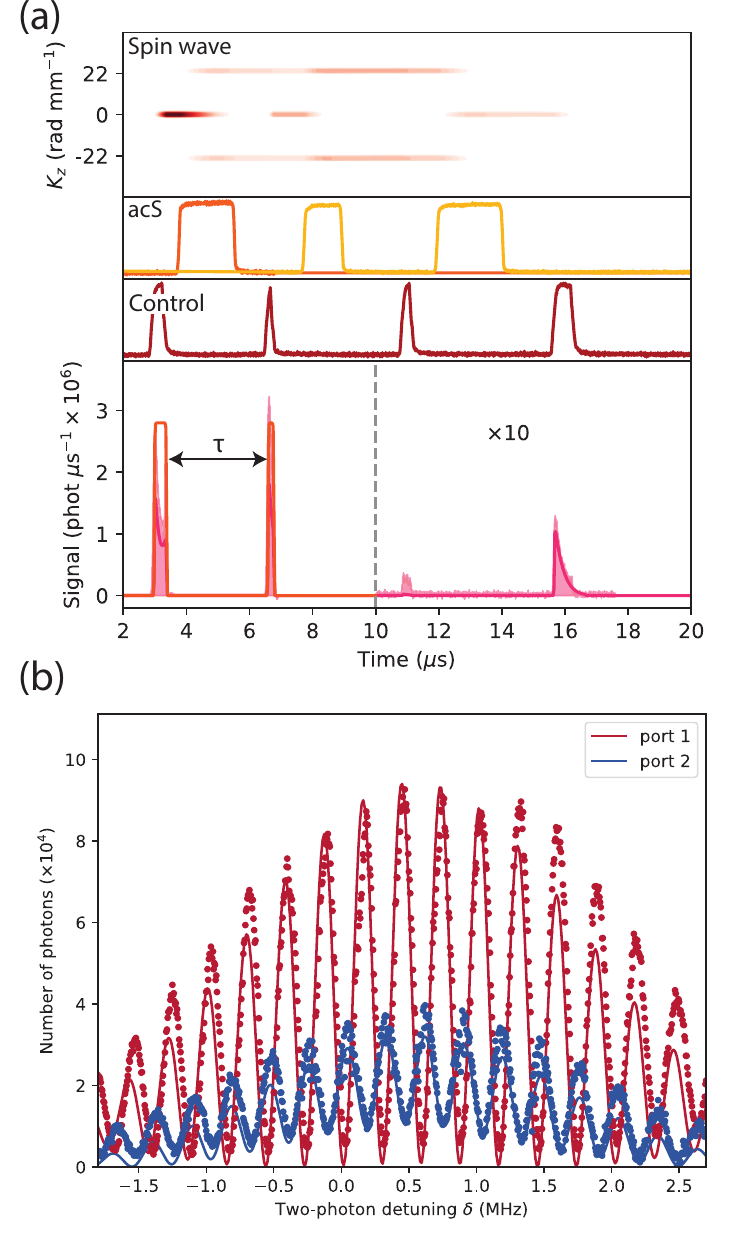}

\caption{Mach-Zehnder interferometer for pulses stored at different times.
(a) Time- and wavevector-resolved spin-wave density $S(K_{z},t)$
during operation of the beamsplitter for a specific choice of phase
leading to constructive interference in the second port along with
time traces of ac-Stark, coupling and signal fields. (b) Ramsey fringes,
i.e. light intensities registered in two ports of the beamsplitter
as a function of two-photon detuning $\delta$.}
\end{figure}

To demonstrate the beamsplitting capability for pulses arriving at
different times we use again the triangle-wave modulation in the $z$-direction,
with $k_{\tri[0.08]}=22\ \mathrm{mm}^{-1}$. After subsequent storage
of two pulses (which is done the same way as in FIFO and LIFO demonstration
using pattern A) we apply the shifted pattern B for a half of period
$T$, modulating the spin wave with amplitude $\pi$ instead of $2\pi$.
This way the two pulses are combined and $K_{z}=0$ component of resulting
spin wave becomes the first output port of the temporal-mode beamsplitter.
Then, after the first readout, we modulate the unread part again with
pattern B with amplitude $\mathcal{A}\approx2.25\pi$ to transfer
a part of the second port to readable $K_{z}=0$ plane, then the second
readout is performed. Note that it is crucial to always perform the
first readout, as otherwise the unread spin wave will interfere and
spoil the operation of the second output port. It is thus necessary
to simulate the operation of this scheme to a full extent, including
possibly imperfect first readout which can affect the second output
port.

To characterize the two pulses interference we change the relative
phase between the pulses by changing the two-photon detuning $\delta$
to observe intensity fringes. Essentially, the phase difference between
the two interfering spin waves is the product of the two-photon detuning
$\delta$ and the time between two first pulses $\tau$. Furthermore,
as we move outside the two-photon resonance the interaction becomes
inefficient. This behavior constitutes the Ramsey interference. In
Fig. 3(b) we plot the total number of photons collected after the
first (port 1) and second (port 2) readout as a function of the two-photon
detuning $\delta$. The observed behavior is properly predicted by
the simulation described in detail in Appendix A. 

\begin{figure}
\includegraphics[width=1\columnwidth]{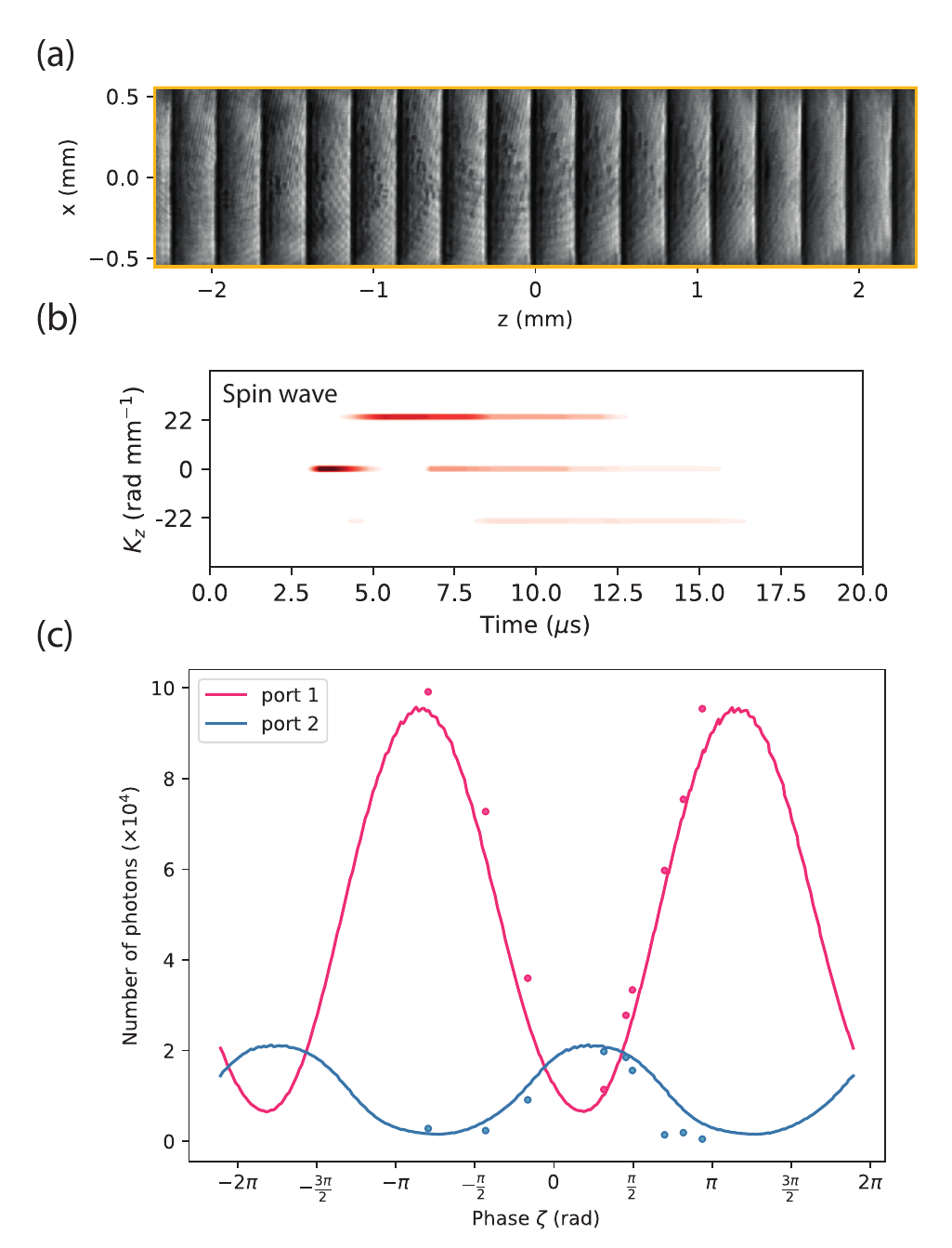}

\caption{Asymmetric interferometer with geometric phase control. (a) The sawtooth-shaped
grating ($\saw$) used to shift the spin waves in one $K_{z}$ direction.
(b) Simulated time and wavevector resolved spin-wave density $S(K_{z},t)$
for a one phase point showing the longitudinal ($K_{z}$) spin wave
mode-mixing during interferometer operation. (c) Intensity interference
fringes for two consecutively retrieved ports (first - port 1, second
- port 2) as a function of sawtooth $\saw$ grating phase $\zeta$.}
\end{figure}

The relative phase between the pulses can be also modified within
the spin-wave domain. To demonstrate this, we implement another interference
protocol; instead of splitting the first pulse into many orders we
simply shift its $K_{z}$ component by $k_{\saw[0.08]}$ using sawtooth
wave $\saw$ modulation 
\begin{equation}
\varphi_{\mathrm{acS}}^{\saw[0.08]}(\xi)=\mathcal{A}^{\saw[0.08]}(\frac{\xi}{2\pi}-\left\lfloor \frac{\xi}{2\pi}\right\rfloor )
\end{equation}
 in the $z$-direction. Then, the second pulse is written to the memory
and the resulting spin wave is modulated using a triangle-shaped grating
of depth $\mathcal{A}^{\tri[0.08]}\approx1.16\pi$ satisfying the
equation $|c_{0}^{\tri[0.08]}|=|c_{1}^{\tri[0.08]}|=|c_{-1}^{\tri[0.08]}|$.
The spatial period of the $\tri$ modulation is chosen to satisfy
$\mathbf{k}_{\tri[0.08]}=\mathbf{k}_{\saw[0.08]}=22\mathrm{\,mm}^{-1}\,\hat{e}_{z}$,
thus the pulses are combined in such a manner that the zeroth order
of the first pulse overlaps with first diffraction order of the second
pulse and conversely. The first interferometer port is again a resulting
$K_{z}=0$ spin wave component so it can be completely readout without
any additional manipulations. The second port this time is well defined
and lies at a plane with $K_{z}=k_{\tri[0.08]}$. Thus, in principle
the second port could be restored completely by applying reversed
sawtooth pattern $\sawr$ shifting back the spin wave by $\mathbf{-k}_{\saw[0.08]}$
to readable region in wavevector space. Due to our setup limitations
(see appendix B) we probe the second port by applying the $\tri$
modulation with amplitude equal $\pi$ and subsequently the phase-matched
component ($K_{z}=0$) is released. The relative phase between interfering
components can be manipulated by changing the phase of one of the
gratings ($\saw$ or $\tri$), as for any shifted periodic modulation
$\varphi_{\mathrm{acS}}^{\mathrm{periodic}}(\xi-\zeta)$ the complex
amplitudes of subsequent orders change as $c_{n}\sim e^{in\zeta}$.
We directly witness this behavior by shifting the sawtooth $\saw$
grating portrayed in Fig 4(a) in the $z$-direction and measuring
interference fringes in the total energy of the released pulses. In
Fig. 4(c) we plot the resulting interference pattern, accompanied
with a proper simulation, showing the interference in wavevector space
{[}Fig. 4(b){]}.

\section{Transverse space interference and manipulation\label{sec:Transverse-space-interference}}

\begin{figure}
\includegraphics[width=1\columnwidth]{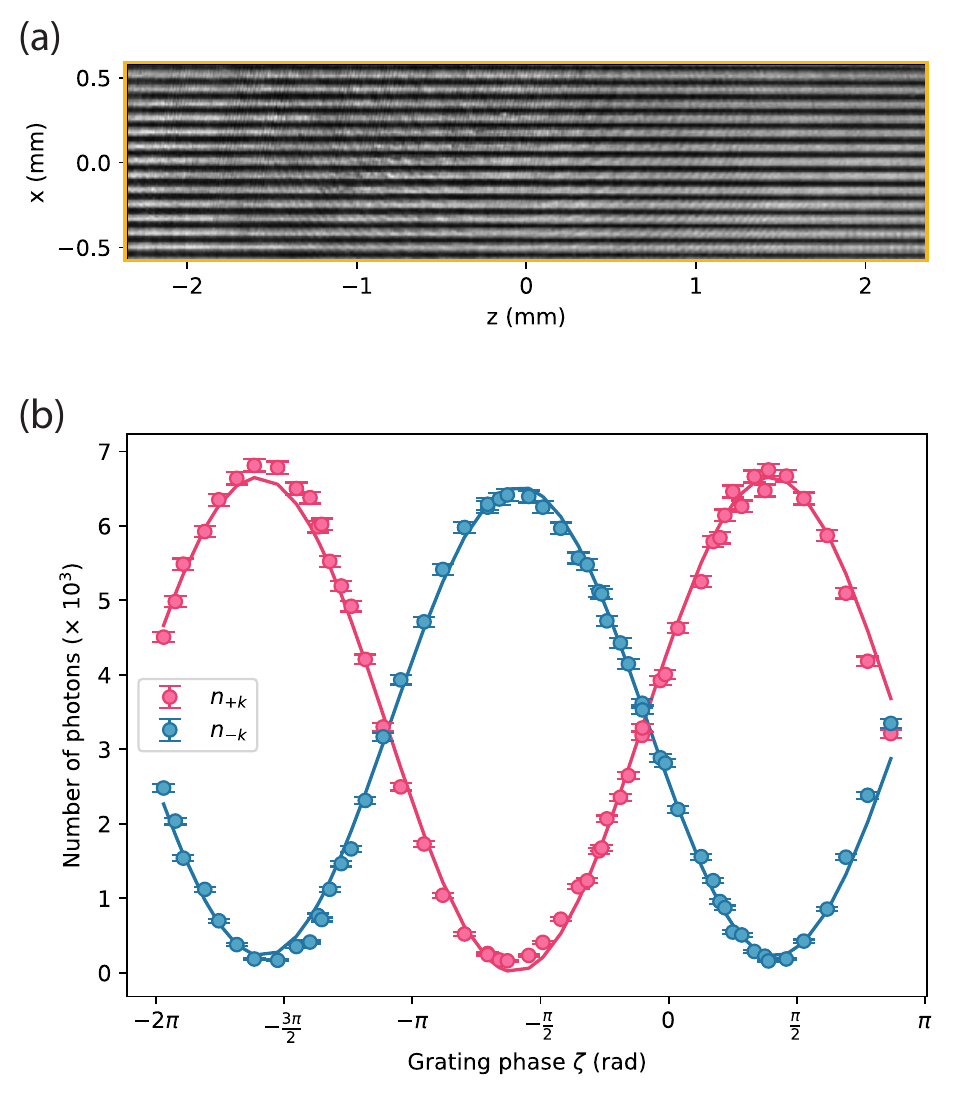}\caption{Spin-wave interference induced by ac-Stark shift manipulation in the
transverse-wavevector space. (a) The sinusoidally-shaped $\sim$ grating
pattern used to apply transverse-mode ($K_{x}$) beamsplitter transformation
by ac-Stark modulating the spin waves. (b) Intensity interference
fringes for two transverse-mode ports separated by $2k$ detected
on the sCMOS camera situated in the far-field of the ensemble. The
grating phase $\zeta$ is changed using piezo-actuated mirror. \label{fig:Spin-wave-interference-induced}}
\end{figure}

To go beyond a single transverse mode we now add the $K_{x}$ dimension
to the scheme. In a simple yet highly robust scenario we map two equally
bright pulses arriving at the same time yet into two different spin
waves with $K_{x}=\pm k$, where $k=75.4\ \mathrm{mm}^{-1}$ using
a pattern presented in Fig. \ref{fig:Spin-wave-interference-induced}(a).
We than apply a sinusoidal grating modulation:
\begin{equation}
\varphi_{\mathrm{acS}}^{\sim}(\xi)=\mathcal{A}^{\sim}(\sin(\xi+\zeta)+1)/2
\end{equation}
 with $k=2k\hat{e}_{x}$ and $\mathcal{A}^{\sim}\approx0.46\pi$ that
again satisfies $|c_{0}^{\sim}|=|c_{1}^{\sim}|=|c_{-1}^{\sim}|=\mathcal{{C}}\approx0.55$.
In this way the output ports at $K_{x}=\pm k$ are mixtures of both
input ports in the 50:50 ratio. We than again use the fact that shifting
the grating position $\zeta$ changes the phase at orders $\pm1$
by $\pm\zeta$. We may thus write the (lossy) beamsplitter transformation
as:

\begin{equation}
\left(\begin{array}{c}
\mathcal{{E}}_{+k}^{\mathrm{{out}}}\\
\mathcal{{E}}_{-k}^{\mathrm{{out}}}
\end{array}\right)=\mathcal{{C}}\left(\begin{array}{cc}
1 & e^{i\zeta}\\
e^{-i\zeta} & 1
\end{array}\right)\left(\begin{array}{c}
\mathcal{{E}}_{+k}^{\mathrm{{in}}}\\
\mathcal{{E}}_{-k}^{\mathrm{{in}}}
\end{array}\right)
\end{equation}

We scan the phase using a piezo-acctuated mirror mount in the far
field of the ensemble (see Appendix B for details of the imaging setup)
and observe high-visibility interference fringes. Notably, we obtain
average visibility of 95\% by comparing maximum and minimum intensities
observed at each port, as portrayed in Fig. \ref{fig:Spin-wave-interference-induced}(b).
The spin-wave domain interference presented here is a direct classical
analogue of the Hong-Ou-Mandel interference described in Ref. \citep{Parniak2018}.

\section{Simultaneous spin-wave processing in two dimensions\label{sec:2d}}

\begin{figure*}
\includegraphics[width=1\textwidth]{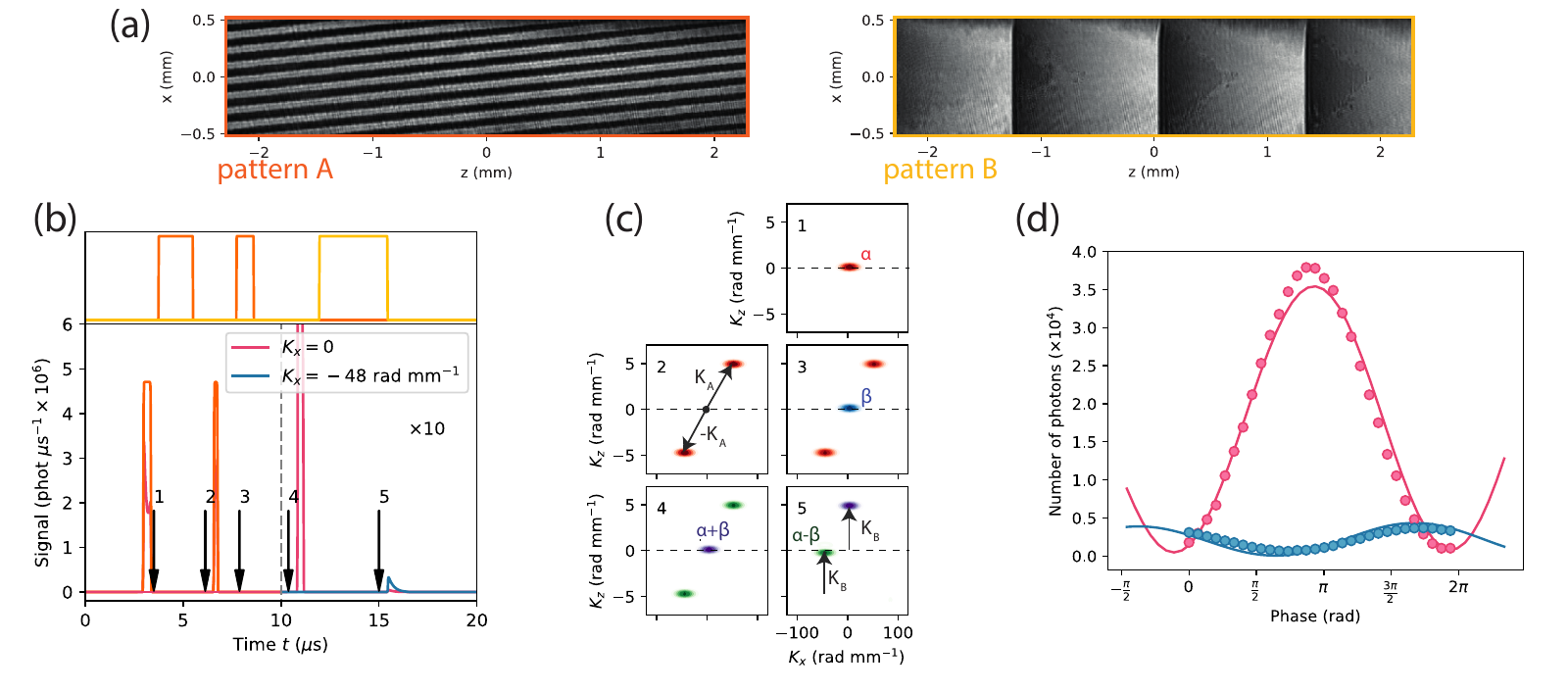}

\caption{Processing and interference in the two-dimensional space. (a) phase
modulation patterns used to split spin waves in the tilted wavevector-space
direction (pattern A) and shift them in the $K_{z}$ direction (pattern
B). (b) Simulated time trace of the scheme demonstrating spin-wave
processing in two wavevector-space dimensions. (c) Simulated wavevector-space
spin wave densities at different time instants of the protocol, with
subsequent numbers corresponding to marked positions in (a). In the
protocol the spin wave with amplitude $\alpha$ is first created by
storing a single coherent pulse (1). Next, the spin wave is split
using pattern A (2) and another pulse is stored to create spin wave
with amplitude $\beta$ (3). Pattern A is then applied again to cause
interference of the two stored pulses (4). Readout is then performed
to first observe the output port corresponding roughly to $\alpha+\beta$,
and after shifting the spin wave in the $K_{z}$ direction with pattern
B the output port corresponding to the $\alpha-\beta$ component.
Panel (d) portrays intensities registered in the two output ports
as a function of relative phase between pulses. Solid lines correspond
to the simulated result.\label{fig:Processing-and-interference}}
\end{figure*}

Finally we combine the longitudinal and transverse manipulations to
exhibit the time-space interference of two sequentially stored pulses.
To access the time-space beamsplitting we design the square-wave grating
${\displaystyle \varphi_{\mathrm{acS}}^{\sq[0.08]}(\xi)=\mathcal{A}^{\sq[0.08]}\left(2\lfloor\frac{\xi}{2\pi}\rfloor-\lfloor\frac{\xi}{\pi}\rfloor+1\right)}$
in the $x-z$ direction $\varphi_{\mathrm{acS}}^{\sq[0.08]}(\text{\ensuremath{\mathbf{k}}}_{\mathrm{acS}}^{\sq[0.08]}\cdot(x,y))$,
where $\text{\ensuremath{\mathbf{k}}}_{\mathrm{acS}}^{\sq[0.08]}=12\ \mathrm{mm}^{-1}\hat{e}_{x}+5\ \mathrm{mm}^{-1}\hat{e}_{z}$.
The periodic collapse-revival behavior of a $c_{0}^{\sq[0.08]}=|\cos(\mathcal{A}^{\sq[0.08]}/2)|$
allows us to use this very grating for both subsequent storage and
interference of two coherent pulses. By applying the square-wave modulation
($\sq$, pattern A in Fig. \ref{fig:Processing-and-interference}a)
of amplitude $\mathcal{A}^{\sq[0.08]}=\pi$ after arrival of the first
pulse and $\mathcal{A}^{\sq[0.08]}=\pi/2$ after the second pulse
is stored we combine the pulses in $t-x$ space. As in previous cases,
the zero-$K_{z}$ component of the processed spin wave becomes the
first port of the Mach-Zehnder interferometer. To sample the second
port (which is spread into successive diffraction orders $c_{n\neq0}^{\sq[0.08]}$)
we use sawtooth grating in the $z$-direction $\varphi_{\mathrm{acS}}^{\saw[0.08]}(k_{\mathrm{acS}}^{\saw[0.08]}z)$
with $k_{\mathrm{acS}}^{\saw[0.08]}=\text{\ensuremath{\mathbf{k}}}_{\mathrm{acS}}^{\sq[0.08]}\cdot\hat{e}_{z}=5\ \mathrm{mm}^{-1}$
(pattern B in \ref{fig:Processing-and-interference}b) to transfer
the 1-st order to readable $K_{z}=0$ space, and perform the readout.
The simulation reveals that in this protocol the two output ports
turn out to not be in perfectly opposite phases. Same behavior is
observed in the experiment, as demonstrated in Fig. \ref{fig:Processing-and-interference}d.
We attribute this effect to imperfect retrieval of the first port
which in turn is partially leaks to the second read-out operation.
We envisage that further simulations will facilitate a more elaborate
scheme that could yield two output ports that are perfectly in opposite
phases, as in experiments described in Sec. V and VI.

\section{Discussion}

We have demonstrated a reprogrammable device that processes atomic
spin waves through interference. Starting with the first demonstration
of an ac-Stark controlled atomic memory for light we have extended
the concept of ac-Stark control to enable interference of coherent
spin-wave states stored in the memory. In particular, the processing
is performed simultaneously in two dimensions of the wavevector space.
With this, we simultaneously exploit temporal and spatial multiplexing.
We show how to perform spin-wave interference between light pulses
stored both at different times, as well as sent to the memory at different
angles. By switching only a pair of patterns we achieve a substantial
degree of reprogrammability and control, which paves the way towards
creating complex unitary quantum networks through spin-wave interference.

The demonstrated optical processor lends itself to many critical schemes
in quantum and classical telecommunication, including the quantum
memory-enabled superaddtitive communication \citep{Guha2011,Klimek2016,Jarzyna2016,Czajkowski2017}
or implementation of a receiver operating with an error rate below
the standard quantum limit \citep{Becerra2013} as well as quantum
metrology through collective measurements on many optical pulses \citep{Demkowicz-Dobrzanski2015,Hou2018}.
The ability of programming interference of stored states provides
a robust tool for probing fundamental properties of quantum systems.
Recently, a tunable beamsplitter transformation have been used to
demonstrate Hong-Ou-Mandel interference between two microwave quantum
memories \citep{Gao2018}. The techniques presented here pave the
way towards programmable complex interference experiments which can
be used to reveal fundamental properties of given quantum system.

The ac-Stark control owes its versatility to the possibly very high
speed of switching and operation, as compared with magnetic field
gradients. This features makes it applicable to recently developed
short-lived quantum memories that operate in the ladder atomic scheme
in warm atomic vapors and achieve very low noise levels \citep{PhysRevA.97.042316,Finkelsteineaap8598}.
The high speed of the ac-Stark control also facilitates real-time
feedback processing that could lead to realization of an even broader
class of operations, including enhanced single-photon generation through
multiplexing \citep{Mazelanik2016,Parniak2017,Nunn2013,Kaneda2017}.
Here, such a scheme could also include engineering of photonic spatial
and temporal mode. This could be taken even further with techniques
used in stationary-light experiments, where amplitude of the stored
spin-wave is non-destructively reshaped using a multi-laser field
\citep{Everett2016,Park2018}. 
\begin{figure*}
\includegraphics[width=0.6\textwidth]{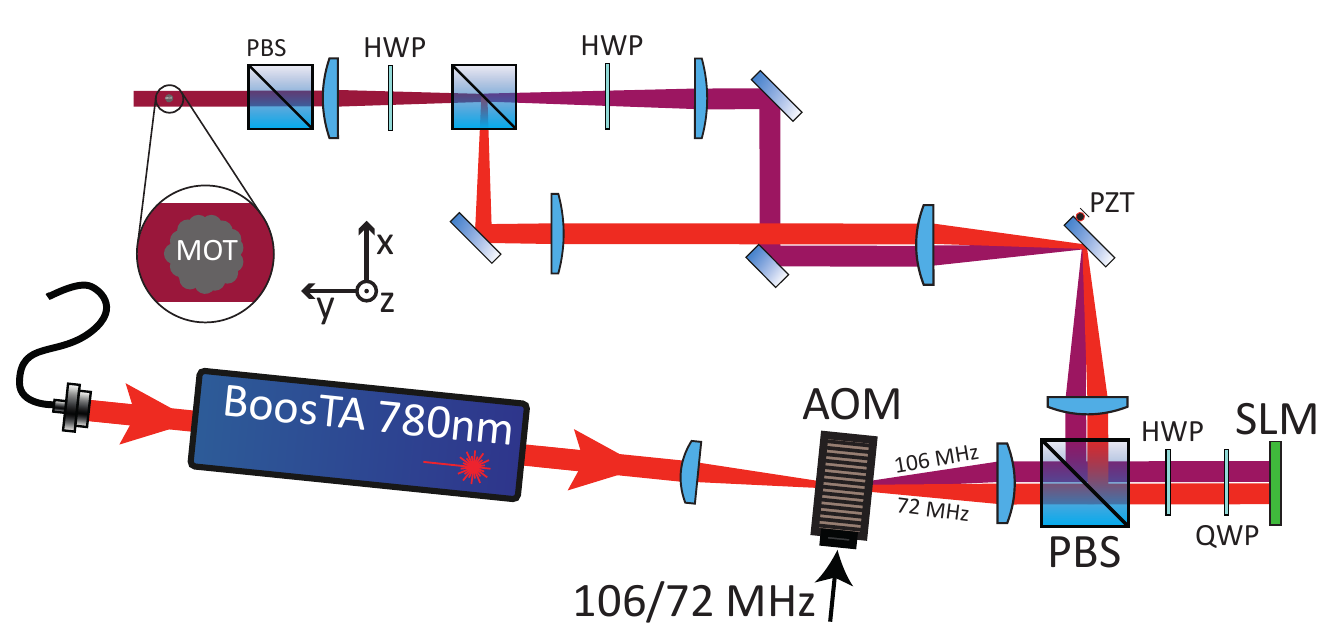}\caption{Schematic illustrating the rapidly-reprogrammable double pattern imaging
system (HWP - half-wave plate, QWP - quarter-wave plate, PZT - piezoelectric
transducer, AOM - acousto-optic modulator). The AOM diffracts the
input beam onto disparate regions of the SLM, which are then simultaneously
imaged onto the MOT after being combined on the polarizing beamsplitter
(PBS). The final PBS projects both light fields onto the $z$ polarization.}
\end{figure*}

Furthermore, note that here we did not use the ac-Stark shift during
write and read operations of the optical memory, and thus the two-photon
absorption line is not broadened. Thus, the Gradient Echo Memory advantage
of avoiding reemision of stored is not yet exploited. Combined with
larger optical densities this could siginficantly improve efficiency
of the presented memory \citep{Sparkes2013}.

Finally, by bringing the presented techniques to spin-waves that involve
a Rydberg state \citep{Firstenberg2013a,Ding2016,Distante2016,Mirgorodskiy2017,Distante2017},
the attainable range of operations between storage modes could be
enriched with nonlinear interactions in order to realize efficient
and deterministic quantum gates for photonic states. This could be
particularly advantageous in engineering complex correlations within
the spatial domain of a Rydberg atomic ensemble \citep{Busche2017}.
\begin{acknowledgments}
We thank K. Banaszek for generous support. This work has been funded
by the National Science Centre, Poland (NCN) (Grants No. 2015/19/N/ST2/01671,
2016/21/B/ST2/02559, 2017/25/N/ST2/01163 and 2017/25/N/ST2/00713)
and by the Polish MNiSW ``Diamentowy Grant'' (Projects No. DI2013
011943 and DI2016 014846).
\end{acknowledgments}

M.M. and M.P. contributed equally to this work.

\section*{Appendix A: light-atom coupling simulations}

To correctly predict efficiencies during storage and retrieval as
well as non-trivial shapes of spin waves created in the atomic ensemble.
We choose to describe the system within the three-level model described
by an interaction picture Hamiltonian within the rotating wave approximation,
which is subsequently reduced using the typical adiabatic elimination
approach by setting the time derivatives of all excited-state coherences
and populations to zero. The coupled equations are then most conveniently
expressed in terms of Rabi frequency of the signal field $\Omega=\mathcal{E}d_{eg}/\hbar$,
where $d_{eg}$ is the dipole moment of the relevant transition, and
the coupling field $\Omega_{C}=\mathcal{E}_{C}d_{eh}/\hbar$. With
the coupling coefficient equal $g=\omega|d_{eg}|^{2}N/2\hbar c\epsilon_{0}$
and in the frame co-moving with the pulses ($t\rightarrow t-z/c$)
the equations take the following form (see Refs. \citep{Parniak2016d,Wasilewski2006,Koodynski2012,Cho2016a}):

\begin{subequations} 
\begin{align}
\frac{{\partial\Omega}}{\partial z}=\  & -ig(N^{-1/2}S^{*}\Omega_{C}+\Omega)/(2\Delta+i\Gamma)\label{eq:deomegapodezet}\\
\begin{split}\frac{{\partial S}}{\partial t}= & \ i\frac{{1}}{2}N^{1/2}\frac{{\Omega_{C}\Omega^{*}}}{2\Delta-i\Gamma}+\\
 & \frac{-2\Gamma\delta-2\gamma\Delta+i\Gamma\gamma+i|\Omega_{C}|^{2}-4i\delta\Delta}{2\left(2\Delta-i\Gamma\right)}S+i\Delta_{\mathrm{acS}}S
\end{split}
\label{eq:deespodete}
\end{align}
\end{subequations}

where we have also introduced $\Gamma=2\pi\times6$ MHz as the excited
state $|e\rangle$ decay rate and $\gamma\approx2\pi\times10$ kHz
as the intrinsic spin-wave decoherence rate, dominated by motional
dephasing. The one-photon $\Delta=2\pi\times20$ MHz and two-photon
$\delta$ detunings are defined as in Fig. 1(b). In the first equation
the two terms in the nominator correspond to the two-photon and one-photon
processes, respectively. In the second equation the first term corresponds
to the Raman interaction, while the second term is the free evolution
under the ac-Stark shift Hamiltonian due to the coupling light, which
includes both the additional phase acquired as well as the power broadening.
Even though the ac-Stark modulation is applied only during dark periods
of the memory, for completeness we include its influence as an additional
term in Eq. \ref{eq:deespodete} given by $i\Delta_{\mathrm{acS}}S$.
Note that in all cases the atom number density $N$ is implicitly
$z$-dependent, and so is the coupling constant $g$. In the simulation
we model this dependence as a Gaussian function in the $z$ dimension
with a width of $5\ \mathrm{mm}$. Finally, we also add a small imaginary
component to the ac-Stark shift $\Delta_{\mathrm{acS}}\rightarrow(1+i\mathrm{{sgn}(\Delta_{\mathrm{{acS}}})}\gamma_{\mathrm{acS}})\Delta_{\mathrm{acS}}$
with $\gamma_{\mathrm{acS}}\sim0.1$, which effectively simulates
dephasing due to inhomogeneous ac-Stark light intensity.

We determine the coupling constant $g$ by observing single-photon
off-resonant absorption. This allows us to experimentally determine
its peak value as $g_{0}\approx200\ \mathrm{{cm}^{-1}}\mu\mathrm{{s}}^{-1}$,
which corresponds to the optical depth $\mathrm{OD}\approx70$. For
the coupling field we take short pulses with smooth slopes (modeling
$\sim100$ ns experimental rise times) and peak $\Omega_{C}\approx2\pi\times9\ \mathrm{MHz}=1.5\Gamma$.
Typical signal field intensities correspond to peak $\Omega\approx2\pi\times50$
kHz. The evolution is simulated using the XMDS package \citep{Dennis2013}
on a two-dimensional $z$-$t$ grid, or three dimensional $x$-$z$-$t$
grid for the results in Section \ref{sec:2d}. For this case we also
include a diffraction term in Eq. \ref{eq:deomegapodezet}, although
the diffraction effects prove to be negligible for the plane-wave
modes we work with . 

\section*{Appendix B: pattern preparation and imaging}

The ac-Stark laser is frequency-stabilized using a offset beat-note
lock \citep{Lipka2017}. It is then spatially filtered using a single-mode
fiber and amplified using a tapered amplifier (Toptica, BoosTA) to
1.5 W. The output beam is then reshaped using a cylindrical lens to
better fit the shape of the elongated atomic ensemble. Simultaneously,
an acousto-optical modulator (AOM) situated in the far field of the
spatial light modulator (SLM) is used to control the position of the
beam at the SLM and carve out $\sim2\ \mu\mathrm{s}$ long pulses.
With this setup we may select which region of the SLM is illuminated
by changing the frequency of the AOM, which is done in real using
a direct digital synthesizer (DDS). On the SLM matrix we display two
patterns in two disparate regions. The SLM surface is then imaged
onto a D-shape mirror which sends each pattern on a different path.
The two paths are joined on a polarizing beamsplitter (PBS) before
the vacuum chamber, but now the two patterns overlap. Note that in
the current configuration we loose half of the power at the final
PBS. An additional mirror placed in the far-field of the SLM is used
for fine adjustment of grating position in the vertical direction
with the help of a piezoelectric transducer (PZT), which is used to
scan the grating phase $\zeta$ as in Sec. \ref{sec:Transverse-space-interference}.

The patterns can also be observed on a camera situated at the same
image plane as the atomic ensemble. The camera provides feedback to
the computer program that controls the SLM, which is used to actually
generate the desired pattern, with particular focus on intensity homogeneity.
The program operates by first mapping the SLM coordinates onto the
camera pixels using a National Instruments Vision module, and then
iteratively adjusting the SLM display to achieve an intensity distribution
closest to the target one.

We envisage that the setup may be extended to feature more patterns
that could be rapidly reprogrammed, by for example using an two-dimensional
AOM to scan the beam through the atomic ensemble, or by using a set
of AOMs to transfer many multiplexed images displayed with an SLM.

\bibliographystyle{apsrev4-1_prx}
\bibliography{../bibliografia}

\end{document}